\documentclass[twoside,fleqn]{article}

\usepackage{espcrc2}
\usepackage[final,dvips]{graphicx}

\title{\bf Future Solar Neutrino Projects}

\author{Robert E. Lanou, Jr.\thanks{This work supported in part
by DoE DE-FG02-88ER40452 \& NSF PHY-9420744.}
  \\ Department of Physics, Brown
University, Providence, RI 02912}

\begin{document}

 \begin{abstract} In the program to determine the intrinsic
properties of the neutrinos and their role in physics and
astrophysics, solar neutrino experiments are playing a fundamental
part. The ``first" and ``second" generation detectors in the field
have been enormously successful and promise to provide still greater
contributions. The challenge to create a ``third generation" of solar
neutrino detectors arises principally from interest in making detailed
investigations of the portion of the neutrino spectrum from a few keV
to 1.5 MeV. That portion contains fluxes from the p-p and CNO
continuum as well as the $^{7}$Be
 and p-e-p lines. The need to overcome
the experimental difficulties presented by working at these low
energies have given rise to new ideas for detector technologies. The
range of technologies is impressive in its variety and reflects
choices of emphasis using  real-time and radiochemical methods. In
this paper, I present a review of the
status of several R \& D efforts known to me which are making progress
toward providing detectors suitable to meet the various challenges of
this low energy region.
 \end{abstract}
\maketitle

\section{INTRODUCTION} The difficulties of creating
effective detectors for solar neutrino research are well known
involving, in varying degree, radioactive backgrounds in the target
and its surroundings, extraction of signal from massive targets,
energy resolution and choices of detection reaction mechanisms.  As lower energy neutrinos are to be detected, most of these difficulties increase or choices are narrowed especially for real-time detectors,
although the higher flux ameliorates somewhat the massiveness of the
targets. As the field of solar neutrino research has matured and moves
toward its goal of achieving a measurement of the shape and
neutrino-flavor composition of the flux over the full spectrum, the
need for a ``third generation" of detectors able to carry out detailed
experiments in the region from a few keV to $\sim1.5$ MeV has been
apparent. This is the region containing the p-p and CNO continuum as
well as the $^{7}$Be and p-e-p lines. From the ``first generation"
radiochemical experiments (Homestake, GALLEX and SAGE) \cite{fi:first} we already know some important things about this
energy region. For example, that the integrated fluxes of $\nu_e$ are
$\leq\frac{1}{2}$ of those expected from the standard solar model
(SSM) \cite{bp:bp85} and that, when taken together in a global analysis
of all solar experiments, suggest that the $^{7}$Be $\nu_e$ flux may be much diminished. From the present status it is not possible to know  how
this flux reduction is shared among the different contributing fusion
reactions nor whether there are neutrino flavors other then $\nu_e$
present.  The upcoming, ``second generation" BOREXINO \cite{bo:borex}
liquid scintillator experiment will directly address the 
$^{7}$Be question in real-time via the (nearly) flavor-independent elastic
scattering reaction on electrons. This represents an important next
step; however, many questions will still remain, among them whether or
not the full p-p flux is seen as $\nu_e$, to what extent is the CNO
flux affected, what is the flavor composition of the flux at all
energies and whether the p-e-p and $^{7}$Be lines can be resolved. The
experimental technology to attack these questions in this energy
regime has not existed but in recent years there have emerged --- and
there are still emerging --- some new approaches in various stages of
research and development which are showing significant promise. No
single technique is likely to emerge which will be able to address all
questions in a single detector and the diversity of the projects I
will discuss reflect choices, on the experimenters' part, to focus on
the different physics opportunities.  In this talk I will try more
strongly to emphasize how some of these new techniques work, some of their recent R \& D results, what still
needs to be done and what their physics foci are rather than what a
finished detector might look like. By and large they are still R \& D
projects and have not yet shown full feasibility for implementation as
a neutrino detector. Some of the projects are known explicitly by
their target material such as Lithium or Gallium Arsenide and others
with the target identity subsumed into acronyms such as HERON, HELLAZ,
LENCSE and GNO.

 \section{HELIUM AS A TARGET} These projects, HERON and HELLAZ, are
both based upon the use of helium as the target medium but they have
radically different approaches and somewhat complementary goals. Both
will utilize the elastic reaction, $ \nu_{e,\mu,\tau} + e^-
\rightarrow \nu_{e,\mu,\tau} + e^-$, for real-time detection in the
energy region dominated by the p-p and $^{7}$Be neutrinos. They will
both measure the energy of the recoil electron and the overall
rate. The similarities and contrasting differences are best seen by
looking at the specifics of each project in turn.\subsection{\bf
HERON:} 
\begin{figure}
\centering 
\setlength{\abovecaptionskip}{1pt}
\setlength{\belowcaptionskip}{1pt}
\includegraphics[width=\columnwidth]{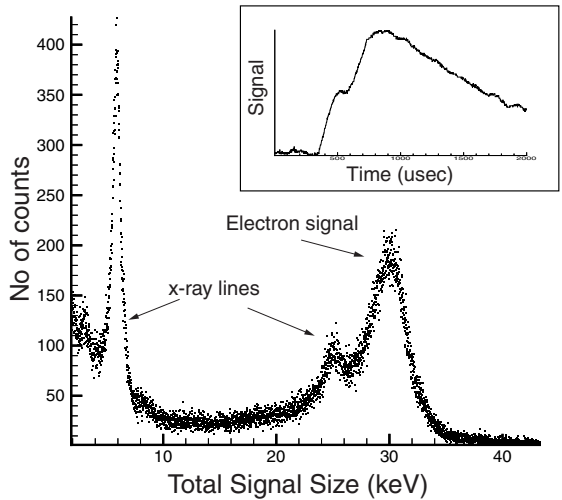}
\caption{Energy spectrum of electrons from $^{113}$Sn
in superfluid $^{4}$He. (Inset: Pulse for single, 364 keV electron
with structure due to early arrival of uv photons.} 
\label{spectrum_fig}
\end{figure}

\begin{figure}
\centering 
\includegraphics[width=\columnwidth]{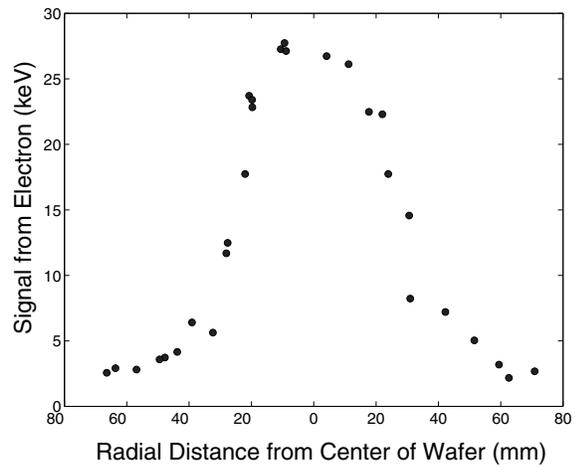}
\caption{Horizontal distribution of electron signal at
2 mm above superfluid surface.} 
\label{angle_fig}
\end{figure}

This project uses $^{4}$He in its superfluid state
\cite{he:heron} and is a new particle detection technique. For purity
from radioactivity, superfluid helium $^{4}$He is ideal for several
reasons. It has no long-lived isotopes, its first excited nuclear
state is at 20 MeV, it is self-cleaning in that all other atomic
species freeze out and $^{3}$He (which has a large neutron cross
section) is easily reduced to negligible amounts in the superfluid during the detector
filling. It has a high density of 0.14 g/cc, it is inexpensive and
standard commercial methods exist for handling the liquid in large
volumes. The mechanism for event identification and energy extraction
from the target utilizes the high multiplicity of carriers ($\sim10^7$ rotons/phonons and $\sim10^3$ uv photons for a 50 keV $e^-$) generated in the liquid by the recoil. An array of low mass silicon
or sapphire wafer calorimeters external to the liquid serve to detect
pulses created by the direct uv photons as well as the delayed pulse
generated by the rotons/phonons through quantum evaporation at the
free surface. Event position in the detector would be found by
observing the spatial distribution of hit wafers and relative times of
arrival of the photon and evaporated atom pulses. The recoil electron
energy would be found by the total pulse height on all hit
wafers. Backgrounds are primarily Compton electrons from residual
activity in the cryostat external to the liquid and would be
subtracted by topological cuts and reference to fiducial and non-fiducial
volumes. A detector large enough to detect 18 p-p and 7 $^{7}$Be
events/day (SSM) would have 10 tons of helium in its fiducial volume
and measure $\sim$5x5x5 meters overall. There would be $\sim1000$
wafer readout channels and both liquid and wafers held at
30mK. Refrigeration on this scale is commensurate with that for
cryogenic gravity wave detectors. \\Through earlier experiments
detecting 3 to 5 MeV alpha particles the group had established the
validity of the basic physics principles of the particle detection
method in superfluid helium combined with the use of wafer
calorimeters. Among other things it was seen that both the prompt uv
photons and delayed roton/phonon evaporation signal were easily
detected on the same silicon or sapphire wafer. Many of the details of
the evaporation process were established including the existence of
the so-called critical angle ($\pm 17^{o}$ for alphas) within which
rotons/phonons were able to initiate evaporation. It was expected that
this critical angle could play a useful role in event position
location.  Nonetheless, it still needed to be established that low
energy electrons could also be detected and with what roton and photon
characteristics. Improvements in wafer sensitivity by adoption of
superconducting thin film thermometers and SQUID technology have been
made to do this. A threshold of 300 eV was reached for energy deposition into an individual wafer. In a recent series of experiments, 364 keV single,
mono-energetic electrons (from $^{113}$Sn radioactive sources movable
in the test cell) have been successfully detected in 3 liters of
superfluid. The measured spectrum is shown in Figure
\ref{spectrum_fig}; the peak at 31 keV is for the electron in this
particular experimental geometry. For calibration purposes x-rays of
5.9 and 25 keV from $^{55}$Fe and $^{113}$Sn sources, respectively,
are deposited directly into the wafer and can also be seen.
As the electron source is moved horizontally below a single wafer the
effect of the critical angle can be seen (Figure \ref{angle_fig}) in the
position dependence of the combined evaporation and photon signal and
suggests a larger critical angle ($\geq30^o$) for the electrons.
Particular attention was paid to studying the total amount of energy
detectable and the resulting partition of that energy between the
evaporation signal and the uv photon signal. In the inset of Figure
\ref{spectrum_fig}, the leading edge shows the prompt photon signal
followed 250 $\mu$s later by the evaporation signal. While both
alphas and electrons yield roughly the same total energy fraction ($\sim10 \% $) they share that energy quite differently between uv
radiation and evaporation. The electron signal relative to the alphas
is nearly 3 times that in the uv photons and about $\frac{1}{2}$ for
the evaporation pulse. This corresponds, for an electron, to 14
photons/keV and 2x10$^5$ (rotons/phonons)/keV . The uv photons have a
wavelength of $\sim80$ nm; a wavelength for which helium is
self-transparent. The group believes it has completed its study of the
physics processes underlying the proposed particle detection technique and has a good understanding of the mechanisms involved.
The results are consistent with a microscopic model in which the
initial ionization density along the track governs the rate of
formation and collision relaxation of helium dimers and therefore the relative portions of energy detectable as rotons/phonons and uv photons. \\Among the
future work toward demonstrating feasibility for solar neutrino
detection are several items. Foremost is the need to improve the
sensitivity of the wafers for use in large volumes by a factor
$\sim20$ . This is closely related, in part, to trying to capitalize
on the increased uv radiation found to be emitted by the low energy
electrons and incorporating it most effectively with the evaporation
signal. A particularly promising avenue being pursued for increased
sensitivity is based on incorporating new developments in magnetic
calorimetry. Further work with superconducting thin film thermometers is
also in progress. A multi-wafer cell is under development to test
various features of energy and position determination as well as
implementation of cryogenics for such a system.
\subsection{\bf HELLAZ:} This project \cite{hl:hellaz} will utilize helium (mixed with $\sim1 \%$ CH$_{4}$) in gaseous form under pressure and reduced temperature (5 atmos. and 77K)
yielding a 0.003 g/cc density. It seeks to use an existing particle
detection technique and scale it up in size (2000 $m^{3}$) and into
the high pressure and low temperature regime. The mechanism for event
identification and energy extraction makes use of the drift of the
ionization electrons ($\sim$4x10$^3$ for 100 keV recoil $e^-$) in a
large time projection chamber (TPC). The event signature consists of
full $e^-$ recoil track reconstruction. The recoil energy is to be
determined by counting individual secondary, ionization electrons as
they drift to the x-y imaging plane. Use of the first arriving
secondary electron to set a $t_0$ and the time delays of the
subsequent secondaries on the earliest part of the track should allow
determination of the track orientation in the TPC. Detection of the Bragg
peak should distinguish front from rear track end. Combination of this
information with the contemporaneous position of the Sun would be used
to deduce the energy of the incident neutrino. Backgrounds here are of
two types: a) Compton electrons from the TPC and the pressure vessel
and b) from beta decay of $^{14}$C in the gas mixture. Backgrounds
would be subtracted using a sample of events constructed assuming the
Sun to be located $180^o$ from its true position. A detector of 6 tons
would be expected to yield 11 p-p and 4 $^{7}$Be events/day (SSM); it
would be 25 meters long and 10 meters in diameter. With x-y detector
planes at both ends the recoil electron track would be drifted up to
10 meters. If experimental hall dimensions require, it could in
principle by split into two portions. \\In earlier work the group had
extensively tested various gas mixtures to find one suitable for TPC
operation under the conditions envisioned for the solar neutrino
application. 
\begin{figure}
\centering 
\fbox{\includegraphics[width=\columnwidth]{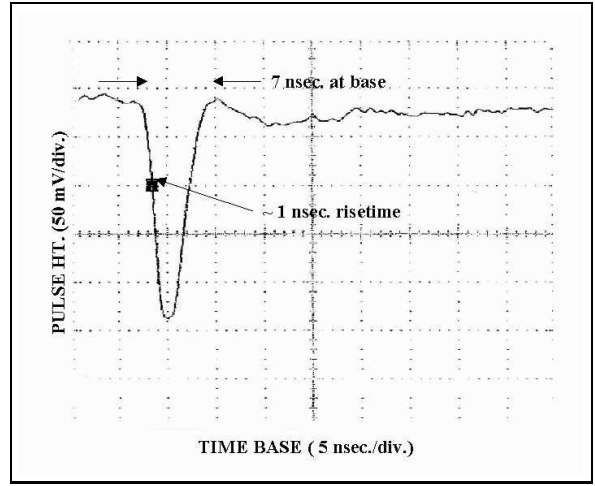}}
\caption{Single photoelectron pulse in ``micromegas" chamber.}
\label{single_pulse_fig}
\end{figure} 
\begin{figure}
\centering 
\fbox{\includegraphics[width=\columnwidth]{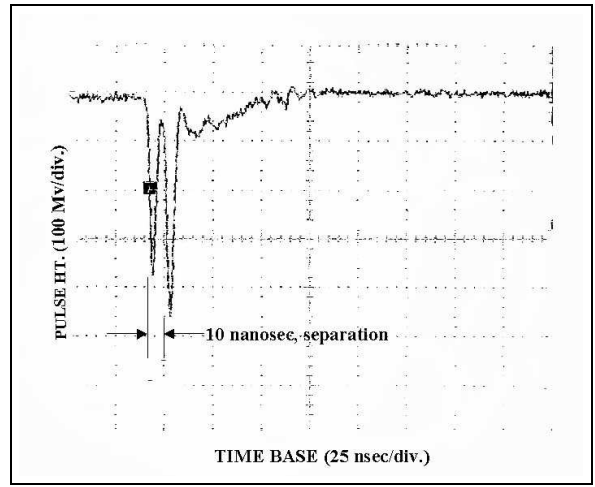}}
\caption{Two, successive photoelectron pulses in
``micromegas" chamber.}
\label{double_pulse_fig}
\end{figure}
The He-CH$_4$ mixture was selected as the most promising but it
imposes severe requirements on the speed ($\leq20$ ns) and gain (
$\sim10^{6}$) needed in the x-y readout for single secondary electron
counting which is to provide the energy and space orientation of the
recoil track. However, recently they have made important progress on
that front by testing a new wire chamber technology --- the
``Micromegas" \cite{mi:micro} chamber. In that test a small chamber
was used which was filled with a 1 atmosphere helium and 6\%
isobutane mixture and constructed of fine mesh forming a 100 $\mu$m
gap above a one-dimensional strip read. An x-ray source of known
strength produced single photoelectrons in the chamber. These were
detected with a rise time of $\sim1$ ns and a gain of
2x10$^{6}$. Details of a single photoelectron pulse are shown in
Figure \ref{single_pulse_fig}. The successful detection of two,
successive photoelectrons separated by only 10 ns is illustrated in
Figure \ref{double_pulse_fig}.  
\\Future work is planned on several fronts. Of particular
importance is the need to adapt this counting success to the large
scale TPC environment : e.g., extend to two dimensional capability,
find a gas and operational compatibility with the track drift and
manage the potential, large increase in readout channels. Several
ideas for achieving these points are under active consideration among
them, incorporating an electrostatic lens to reduce the effective area
of the 100 $m^{2}$ endcaps. Nearing completion is a 5 liter TPC
capable of 5 bar and 77K operation to be used in tests to produce and
detect Compton recoils from 511 gammas. They are also preparing to
initiate tests for low activity materials as well as to carry out
further Monte Carlo studies related to recognition of electron track
direction and energy determination. \section{\bf LENCSE:} A recent
proposal \cite{yb:ybex} has been made to utilize $^{176}$Yb in a
real-time, liquid scintillator detector. By detecting $\nu_e$ through
inverse beta decay it would serve as a complement to those detectors
using the flavor independent elastic scattering from electrons. This
project will be discussed in a separate presentation by
Dr. Raghavan. \section{CRYOGENICS IN RADIOCHEMICAL EXPERIMENTS} A new
approach to counting the back-decay of atoms collected in
radiochemical experiments is being carried out by two groups (GNO and
Lithium). Although this new technology has not yet been perfected for
use in an operating solar neutrino detector, the groups are making
impressive progress.  The idea is to replace the classical method of
small proportional counters by cryogenic
micro-calorimeters. Micro-calorimeters are small (few micro- or
milli-gram) crystals at milli-Kelvin temperatures to which the
extracted radioactive sample has been applied. Because of the
exceptionally low heat capacity of the calorimeter, and if $4\pi$
coverage is achieved, then the Auger electrons and x-rays following
the electron capture decay are contained hermetically and a
temperature pulse results. This is a technique which is developing
rapidly in other areas (e.g., for x-ray astronomy, double beta decay
or dark matter searches ) where resolutions less than 100 eV have been
obtained. Although the primary motivations for the two groups
discussed here to employ the cryogenic method are somewhat different,
the technique holds out promise in several areas important to both:
lower thresholds, improved energy resolution, full energy deposit and
increased counting efficiency. A particular challenge common to both
experiments will be to devise a method which ensures that a fully
efficient transfer of the precious sample, extracted
from the chemical reactor, is made onto the micro-calorimeter
crystal. \subsection{\bf Lithium:} 
\begin{figure}
\centering 
\includegraphics[width=\columnwidth]{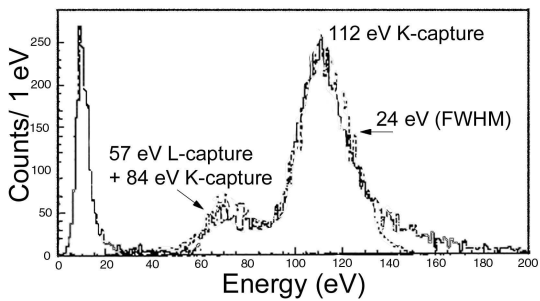}
\caption{Spectrum of ${}^{7}Be \rightarrow {}^{7}Li$
and ${}^{7}Be \rightarrow {}^{7}Li^{*}$ in micro-calorimeter.}
\label{be7_spec_fig}
\end{figure}

For some time, $^7$Li has attracted
interest as a radiochemical target with special relevance for the
p-e-p and CNO neutrinos because the strength of the ground and first
excited states can be accurately inferred from laboratory experiments
\cite{ba:bahcall}. The reaction is $\nu_e$ + $^7$Li $\rightarrow e^-$
+ $^7$Be with a threshold of 862 keV.  A particular challenge to the
realization of a lithium-based detector has been the counting of the
electron capture decays ($\tau_{1/2}$ = 53d) of the extracted $^7$Be
in which 90\% go to the $^7$Li ground state but produce an Auger
electron of only 55 eV together with a 57 eV nuclear recoil and is
therefore not amenable to the usual proportional counter
methods. Consequently, interest had centered on utilizing the decay to
the first excited state by detecting the subsequent 474 keV gamma but,
with only a 10\% branching ratio, a 100 ton detector was required to
achieve a rate of 0.5/day. Presently, a joint Russian-Italian project
\cite{li:lithium} has made interesting progress on two fronts: a)
first, by capitalizing on the use of the 90\% channel by developing
the micro-calorimeter method thus making conceivable a detector of
only 10 tons and b) second, by developing a large scale prototype for
the lithium reactor itself to test the difficulties of the lithium
handling and sample extraction. \\In tests using cryogenic
micro-calorimeter techniques with accelerator produced samples of Be
and BeO, they have detected $^7$Be decays via the summed energy
deposited by the Auger electron (55 eV) and the recoiling nucleus (57
eV) giving a well resolved peak at 112 eV with a 24 eV FWHM at
$\sim80$\% efficiency (Figure \ref{be7_spec_fig}). The microcalorimeter consisted of
a 100x200x200 $\mu$m NTD thermistor at 45mK glued to the few $\mu$g
sample. 

While further work along these lines is needed, this result seems to
establish feasibility of the method. Among the future projects are
plans to study other Be compounds for improved phonon coupling to the
thermistor and minimization of delayed counting effects from
quasi-particle trapping in the crystal (see the peak at very low
energies).  \\ The mechanical work on a prototype reactor vessel to hold 300 kg of
liquid lithium has been completed at INR(Troitsk). The principal
R\&D work to be carried out with the prototype are studies of the
complete lithium handling method and the Be extraction process into
forms most suitable for the micro-calorimetry counting method. In
order to carry out this important series of tests with the prototype,
funds are urgently needed to move it to another laboratory and to
operate it there. Funding has been applied for and a decision enabling
them to go forward is awaited.  \subsection{\bf GNO:} 
\begin{figure}
\centering 
\includegraphics[width=\columnwidth]{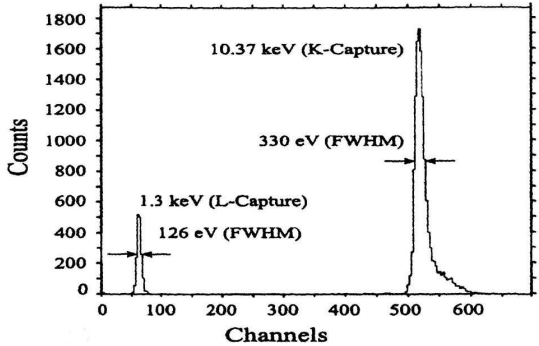}
\caption{Spectrum of ${}^{71}Ge \rightarrow {}^{71}Ga$
in micro-calorimeter.}
\label{ucal_spectrum_fig} 
\end{figure}

\begin{figure}
\centering 
\includegraphics[width=\columnwidth]{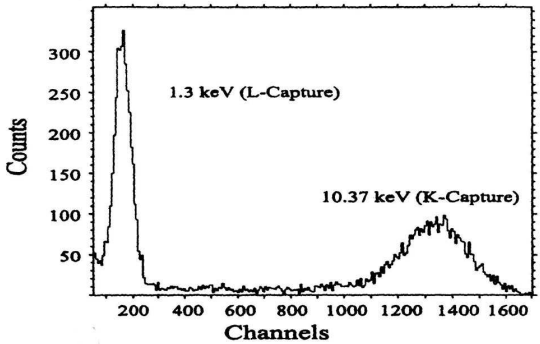}
\caption{Spectrum of ${}^{71}Ge \rightarrow {}^{71}Ga$
in GALLEX proportional counter.} 
\label{GALLEX_spectrum_fig} 
\end{figure}

The Gallium
Neutrino Observatory \cite{gn:gno} is designed to improve and extend
the very successful gallium radiochemical technique. Details of GNO's
goals and present operating status can be found in Dr. Kirsten's
presentation. My emphasis here is on the potential, cryogenic
innovations which may have broad application. The events to be counted
in GNO are the electron capture decays of $^{71}$Ge. In the recently
completed GALLEX experiment, 70\% of the contribution to the
systematic error was from the energy acceptance window. Consequently,
the much improved energy resolution which the micro-calorimeters
promise is one reason arguing for attempting to exploit them. Another
reason is the potential for increased counting efficiency; the
proportional counter efficiencies are typically 70\% due to several
factors including dead volume and missed x-rays. Every improvement in
efficiency allows a similar reduction in gallium mass for the same
statistics. Improved systematics and statistics are among the
principle goals of GNO.  K-capture constitutes 88\% of the decays ($\sim41\%$ each into the 10.37 keV Auger electron or a 9.35 keV
$K_\alpha$ with a 1.12 keV $e^-$ and 5.3\% into a 10.26 keV $K_\beta$
with a 0.11 keV $e^-$). The L-capture is 10.3\% of the rate giving
only a 1.30 keV Auger electron. Hermetic detection of the x-ray would
prevent the ``contamination" of the L-peak by missed $K_\alpha$
x-rays. An attractive bonus of an additional 1.7\% of the rate could
be gained if the threshold can be kept below 150 eV since M-capture
would be accessible. \\The Munich group has made significant progress
in experiments for cryogenic detection of $^{71}$Ge. Two types of
experiments are of particular interest. In the first, 1 $mm^3$ crystal
of $^{nat}$Ge was neutron activated thus producing $^{71}$Ge whose
decays had $4\pi$ coverage. Attached to the crystal was a tungsten
superconducting thin film transition edge thermometer (at 15 mK)
with which the temperature pulses were measured. The resulting
spectrum is shown in Figure \ref{ucal_spectrum_fig} and for comparison
Figure \ref{GALLEX_spectrum_fig} is a typical GALLEX proportional
counter result for the same decay. The effect of $4\pi$ acceptance and
improved resolution is clearly evident.
In a solar neutrino
experiment it is Ge from GeH$_{4}$ which must be attached to a
crystal. As a first step in using GeH$_{4}$ a thermally decomposed
sample of GeH$_{4}$ was deposited onto a 1$\mu$g sapphire crystal
equipped with an iridium-gold transition edge thermometer. Due to the
$2\pi$ nature of the decay coverage the detection was not hermetic;
however good resolution (160 eV on the 5.89 keV Mn $K_\beta$
calibration line) was obtained and the features of the $^{71}$Ge decay were clearly
observed. They have also done some initial work on $^{37}$Ar. \\This
work is continuing and among the immediate projects are improvements
in the electronics to obtain a threshold low enough for seeing the
M-line, developing a two crystal ``sandwich" as a $4\pi$ hermetic
device with readout on both sides for energy summing, reducing the
contributions from deposition in the thermometer glue (see the
high-side tail in Fig. 6 ) and, crucially, finding a high efficiency
method for placing the Ge from GeH$_{4}$ on the crystals. \section{\bf
GALLIUM ARSENIDE:} This is perhaps the most ambitious project
\cite{bw:bowles} and while it does not aim to be the ``complete" solar
neutrino detector it does give a measure of the task to realize a
single detector more inclusive of reaction types and range of
energies. Its principal aims are to make a model independent test of
flavor non-conservation (including sterile $\nu$'s), to determine
the precise energies and widths of the lines of $^7$Be (862 keV),
p-e-p (1.4 MeV) and the end point of the p-p continuum using the
three channels of elastic, charged current and neutral current
scattering of $\nu$'s. Further, in the case of any observed flavor
non-conservation, they wish to determine the MSW parameters precisely
if that is the mechanism involved. To carry out such a program, a
massive detector with $\leq 2$ keV energy resolution is
required. Their idea is to create a GaAs based, electronic, real-time
detector. Gallium and arsenic are chosen because they have no
long-lived isotopes or (n,$\gamma$) daughters and for the possibility
of good resolution due to the high multiplicity of e-hole pairs and
low noise when cooled. However, they estimate this would require a 125
tonne (60 t of Ga) in 40,000 hyperpure 3.2 kg segments. Among the
primary reactions to be exploited are $ \nu_{e,\mu,\tau} + e^-
\rightarrow \nu_{e,\mu,\tau} + e^-$, $\nu_e + {}^{71}Ga \rightarrow
e^- + {}^{71}Ge$ and $ \nu_{e,\mu,\tau} + {}^{71}Ga \rightarrow
\nu_{e,\mu,\tau} + ({}^{71}Ga)^*$. Good resolution of the p-e-p and
upper $^7$Be line would be an important advance; however, a successful
data analysis to achieve fully the above goals will be a very
challenging one in that it relies upon use of the shape of the elastic
differential cross-section to separate $\nu_e$ and $\nu_{\mu,\tau}$ as
well as requiring a separation of the charged current p-p
events. Extreme purity (from U, Th, $^{40}$K \& $^{14}$C) of the
GaAs as well as its electronic performance for devices as large as 3.2kg
must be tested and assured. \\ R\&D on this project has recently
begun. It centers primarily on three areas: a) production of testable,
small GaAs devices made from sizable boules, b) tests of the
electronic properties of these small devices and c), based on results
from these tests, investigation of new crystal growing methods to
produce improved quality and larger boules.  They report that in
Russia twenty, high quality 1 kg GaAs ingots have been grown for
them. And that from these ingots, several 400 $\mu$m thick working
detectors have been made and their electrical properties
measured. They find that most properties for these small devices are
equal in quality to the best in the field but still fall short of what
would be needed in the full-sized solar neutrino device. They find
high electron mobility but need significant improvement in charge
collection. They are preparing to take the next step to increase the
thickness of the electrical test devices to 1 mm thickness and they
are studying new crystal growing methods adaptable to their existing
furnaces. \\ There is still a very, very long way to go on this very
bold project but if it can be shown to be feasible it may be the
closest we will ever get to a ``universal" solar neutrino
detector.\section{COMMENTS \& CONCLUSIONS:} The field of solar
neutrino physics has certainly come of age as the exciting results so
far are showing. From Super-Kamiokande and the other second generation
experiments, SNO, BOREXINO and GNO, just coming on line we can expect
new crucial tests of neutrino properties. However, it is well to keep
in mind that these experiments and their predecessors would not have
been possible without the invention of new (or by pushing to the
limit) experimental techniques. We are still only part of the way to
having the experimental capabilities we need over the full solar
spectrum. In order to go beyond what we now can do, we need new
techniques. \\ I have tried to show by the projects described here
that there is an active world-wide R \& D effort making important
progress on several fronts toward these goals; but we should keep in
mind that we are not fully there yet and there is always room for
more, new ideas. \section{ACKNOWLEDGMENTS:} I am very grateful to J. Adams, M. Altmann, T. Bowles, V. Gavrin, P. Gorodetzky, T. Kirsten, A. Kopylov, R. Raghavan, S. Vitale, and T. Ypsilantis for very helpful information and discussion concerning the specifics of the projects in which they are involved.

\end{document}